\documentclass[aps,prb,superscriptaddress,twocolumn,showkeys]{revtex4}

\usepackage{graphicx}

\begin{document}

\title{Magnetocaloric effect and magnetization in a
Ni-Mn-Ga Heusler alloy in the vicinity of magnetostructural
transition}

\author{A.~Aliev}
\author{A.~Batdalov}
\affiliation{Institute of Physics of Dagestan SC RAS, Makhachkala
367003, Russia}

\author{S.~Bosko}
\author{V.~Buchelnikov}
\affiliation{Chelyabinsk State University, Chelyabinsk 454021,
Russia}

\author{I.~Dikshtein}
\affiliation{Institute of Radioengineering and Electronics of RAS,
Moscow 101999, Russia}

\author{V.~Khovailo}
\affiliation{National Institute of Advanced Industrial Science and
Technology, Tohoku Center, Sendai 983-8551, Japan}

\author{V.~Koledov}
\affiliation{Institute of Radioengineering and Electronics of RAS,
Moscow 101999, Russia}

\author{R.~Levitin}
\affiliation{Physics Depertment, Moscow State University, Moscow
119899, Russia}

\author{V.~Shavrov}
\affiliation{Institute of Radioengineering and Electronics of RAS,
Moscow 101999, Russia}

\author{T.~Takagi}
\affiliation{Institute of Fluid Science, Tohoku University, Sendai
980-8577, Japan}

\begin{abstract}
The magnetic and thermodynamic properties of a
Ni$_{2.19}$Mn$_{0.81}$Ga alloy with coupled magnetic and
structural (martensitic) phase transitions were studied
experimentally and theoretically. The magnetocaloric effect was
measured by a direct method in magnetic fields $0-26$~kOe at
temperatures close to the magnetostructural transition
temperature. For theoretical description of the alloy properties
near the magnetostructural transition a statistical model is
suggested, that takes into account the coexistence of martensite
and austenite domains in the vicinity of martensite transformation
point.
\end{abstract}

\keywords{Ferromagnetism; Magnetostructural phase transition;
Magnetization; Magnetocaloric effect}

\maketitle

In recent year Ni$_2$MnGa-based ferromagnetic shape memory alloys
have attracted considerable attention as a new class of actuator
materials (see, for a recent review, Ref. [1]). Giant
magnetocaloric effect (MCE) has also been observed in these
alloys. Solid state materials with a large MCE are considered to
be very perspective in creation of solid state refrigerators with
high efficiency, manufacturability and ecological compatibility
(Ref. [2] and references therein). In Ni-Mn-Ga alloys, the
martensitic transition temperature is very sensitive to
composition, which is very important for practical applications of
these alloys. For instance, upon a partial substitution of Mn for
Ni the martensitic transition temperature $T_m$ increases and
Curie temperature $T_C$ decreases until they merge in one
first-order magnetostructural phase transition (MSPT) [1]. The
values of MCE in the vicinity of such a transition are likely to
be maximal. The aim of our work has been to study, experimentally
and theoretically, magnetic and thermodynamic properties of
Ni$_{2.19}$Mn$_{0.81}$Ga alloy with MSPT and to determine directly
the magnitude of MCE.

Polycrystalline Ni$_{2.19}$Mn$_{0.81}$Ga alloy was prepared by a
conventional arc-melting method in argon atmosphere. MCE was
measured by a direct method in a calorimetric system at constant
temperatures in the vicinity of MSPT. The sample for the
measurement of MCE was fixed by glass wires of 10~mm in diameter
glued to the sample. The change in the temperature of the sample,
$\Delta T$, with the change in magnetic field, $\Delta H$, was
measured by a differential thermocouple. A cooper screen was
placed near the sample in order to decrease radiational losses.
The magnetic field up to 26~kOe was created by an electromagnet.

The temperature dependence of $\Delta T$ at turning the magnetic
field off is presented in Fig.~1. The experimental results
obtained evidence that the $\Delta T$ dependence has a peak at $T
\approx 340$~K. This temperature corresponds to the temperature of
MSPT [1].

\begin{figure}[h]
\begin{center}
\includegraphics*[width=6cm]{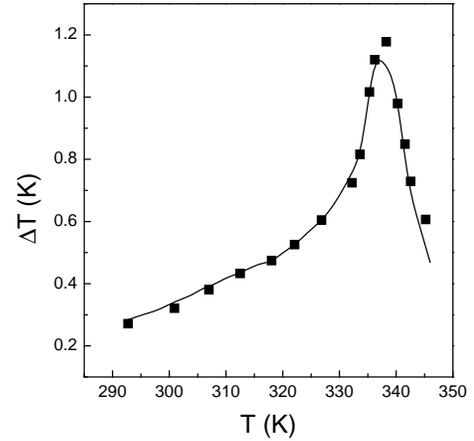}
\end{center}
\caption{Temperature dependence of sample temperature change
$\Delta T$. Points are experimental data, the solid curve is
theoretical one.}
\end{figure}

A statistical model is suggested for the theoretical description
of the alloy properties near MSPT. The model takes into account
coexistence of three types of martensitic and one type of
austenitic structural domains in the vicinity of MSPT point [3,4].
Each structural domain is divided by $180^{\circ}$ magnetic
domains. It is postulated that the rate of transformation between
structural domains is proportional to the net probability of
transformation. This probability is expressed using the value of
energetic barrier between the domains

\begin{equation}
P_{\alpha\beta} = \mathrm{exp}\Big(-\frac{\Delta
Vb_{\alpha\beta}}{kT}\Big)
\end{equation}

\noindent where $\Delta V$ is the transformation volume,
$b_{\alpha\beta}$ is the energetic barrier for transitions from
$\alpha$ phase to $\beta$ phase ($\alpha$ and $\beta$ are the
numbers of structural domains). The magnitude of the potential
barrier can be obtained as the energy at the point where Gibbs
potentials of $\alpha$ phase and $\beta$ phase are equal for a
fixed value of stress. Gibbs potential is $G=F-SE$, where $F$ is
free energy, $S$ and $E$ are stress and strain. The free energy of
the sample consists of magnetic, magnetoelastic and elastic terms.
The evolution of the system depends on the energetic impacts of
each phase. The rate of transformation from one phase to another
is determined by

\begin{equation}
\dot{\xi}^\alpha = \sum_{\alpha}^{\beta\ne\alpha}
\omega(P_{\beta\alpha}\xi^{\beta} - P_{\alpha\beta}\xi^{\alpha}),
\end{equation}

\noindent where $\omega$ is the frequency of  attempt,
$\xi^{\alpha}$ is a volume fraction of $\alpha$ phase. From
equation (2) we can find the volume fraction of each phase. Then
the magnetization of the sample can be found as

\begin{equation}
M = \sum_{\alpha=1}^4 \xi^{\alpha}M^{\alpha},
\end{equation}

\noindent where $M^{\alpha}$ is the magnetization of $\alpha$
phase. Using (3) we can obtain the temperature change
corresponding to the change in magnetic field [5]

\begin{equation}
\Delta T =
-\Big(\frac{T}{C_{P,H}}\Big)\Big(\frac{dM}{dT}\Big)_H\Delta H,
\end{equation}

\noindent where $C_{P,H}$ is specific heat of the alloy.

\begin{figure}[t]
\begin{center}
\includegraphics*[width=6cm]{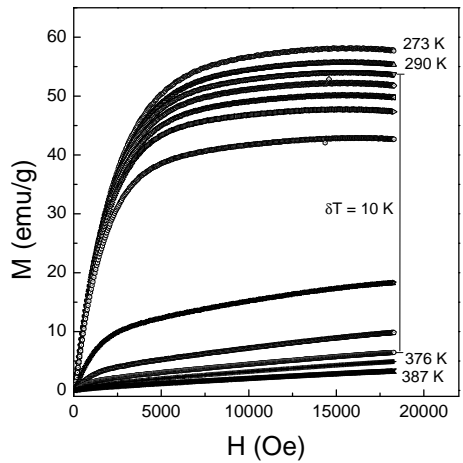}
\end{center}
\caption{Isothermal magnetization curves for
Ni$_{2.19}$Mn$_{0.81}$Ga.}
\end{figure}

Numerical calculation was made with the following parameters of
the alloy $C_{P,H} = 25$~J/kg\,K, $\Delta V =
0.2\times10^{-25}$~m$^3$, $T_m = T_C = 340$~K, $\Delta H =
26$~kOe. The results are presented in Fig.~1 by the solid line. It
is seen that for the given parameters the results of calculation
are in a good agreement with the experimental results.

From Fig.~2 we can determine the magnetic entropy change $\Delta
S$ at MSPT using Maxwell relation. This value is about 4.5~J/kg\,K
at 15~kOe, which is comparable with that observed in the same
magnetic field at the martensitic transition temperature in a
Ni$_{2.104}$Mn$_{0.924}$Ga$_{0.972}$ composition [6].

The work was partially supported by grants RFBR-BRFBR 02-02-81030
Bel2002a, RFBR 03-02-17443, 02-02-16636a and RF Ministry of
Education No E02-3.4-35.


\begin{thebibliography}{20}
\bibitem{1-v} A.N.~Vasil'ev, V.D.~Buchelnikov, T.~Takagi,
V.V.~Khovailo, and E.I.~Estrin, Physics - Uspekhi 173 (2003) 577.

\bibitem{2-t} O.~Tegus, E.~Br\"uck, L.~Zhang, Dagula, K.H.J.~Buschow, and
F.R.~de Boer, Physica B 319 (2002) 174.

\bibitem{3-g} S.~Govindjee and G.J.~Hall, Int. J. Sol. Struct. 37 (2000)
735.

\bibitem{4-b} V.D.~Buchelnikov and S.I.~Bosko, J. Magn. Magn. Mater. 258-259
(2003) 497.

\bibitem{5-v} S.V.~Vonsovsky, \textit{Magnetism} (Nauka, Moscow, 1971), p.402.

\bibitem{6-h} F.~Hu, B.~Shen, J.~Sun and G.~Wu, Phys. Rev. B 64 (2001)
132412.

\end{thebibliography}
\end{document}